\documentclass[aps,prl,reprint,amsmath,amssymb,twocolumn,preprintnumbers,superscriptaddress,10pt,longbibliography]{revtex4-2}

\usepackage[utf8]{inputenc}                                             

\usepackage{xcolor}                                                     

\usepackage{amsmath,amssymb}                                            
\usepackage{amsthm}                                                     
\usepackage[italicdiff]{physics}                                        
\usepackage{bm}                                                         
\usepackage{tensor}                                                     
\usepackage{siunitx}                                                    
\usepackage{simplewick}                                                 

\usepackage{graphicx}                                                   
\usepackage{tikz}                                                       
\usepackage{pgfplots}                                                   
\pgfplotsset{compat=1.14}

\usepackage{hyperref}                                                   

\newcommand*{\complexs}{\mathbb{C}}                                     
\newcommand*{\lp}{\mathopen{}\left}                                     
\newcommand*{\rp}{\right}                                               
\newcommand*{\hc}{\text{h.c.}}                                          

\renewcommand*{\ket}[1]{\left| #1 \right\rangle}                        
\renewcommand*{\bra}[1]{\left\langle #1 \right|}                        
\renewcommand*{\braket}[2]{\left\langle #1 \middle| #2 \right\rangle}   
\newcommand*{\exv}[1]{\left\langle #1 \right\rangle}                    
\newcommand*{\qexv}[2]{\bra{#1} #2 \ket{#1}}                            
\newcommand*{\norder}[1]{:\hspace{1pt}\mathrel{#1}\hspace{1pt}:}        
\newcommand*{\hilbertspace}{\mathcal{H}}                                

\newcommand*{\bignorm}[1]{\big\| #1 \big\|}                             
\newcommand*{\bigabs}[1]{\big| #1 \big|}                                
\newcommand*{\bigket}[1]{\big| #1 \big\rangle}                          
\newcommand*{\bigbraket}[2]{\big\langle #1 \big| #2 \big\rangle}        

\newcommand*{\tbuff}{\mathchoice{\quad}{\>}{\>}{\>}}                    
\newcommand*{\kbuff}{\:}                                                

\newcommand*{\supplementary}{Supplemental Material}                     
\newcommand*{\fref}[1]{Fig.~\ref{#1}}                                   

\hypersetup{%
    pdftitle  = {Approaching single-photon pulses},%
    pdfauthor = {Jan Gulla, Johannes Skaar},%
    colorlinks,%
    linkcolor={blue!50!black},%
    citecolor={blue!50!black},%
    urlcolor={blue!50!black}%
}

\begin{document}


\title{Approaching single-photon pulses}

\author{Jan Gulla}
\affiliation{Department of Technology Systems, University of Oslo, NO-0316 Oslo, Norway}

\author{Johannes Skaar}
\email{johannes.skaar@fys.uio.no}
\affiliation{Department of Physics, University of Oslo, NO-0316 Oslo, Norway}

\date{\today}

\begin{abstract}
Single-photon pulses cannot be generated on demand, due to incompatible requirements of positive frequencies and positive times. Resulting states therefore contain small probabilities for multiphotons. We derive upper and lower bounds for the maximum fidelity of realizable states that approximate single-photon pulses. The bounds have implications for ultrafast optics; the maximum fidelity is low for pulses with few cycles or close to the onset, but increases rapidly as the pulse envelope varies more slowly. We also demonstrate strictly localized states that are close to single photons.
\end{abstract}


\maketitle

It is well known that single photons cannot be strictly localized. The energy density has a tail falling off almost exponentially or more slowly \cite{bialynicki-birula1998}, as dictated by the Paley-Wiener criterion \cite{paley1934,papoulis1962} from the absence of negative frequencies. A photon propagating in the $+x$-direction cannot be localized with respect to $x$ \cite{saari2005} or with respect to time $t$. This means that perfect single photons cannot be generated locally on demand.

On the other hand, classical pulses can be localized to any extent, by having a suitable, broadband spectrum of positive and negative frequencies. Also coherent quantum states can be strictly localized \cite{knight1961,bialynicki-birula1998}. As a result, they can in principle be generated on demand.

Coherent states are however quite different from single-photon states. The natural question is therefore whether it is possible to generate optical states that are close to single photons. Clearly the answer must be yes, as evident from the vast amount of proposals and reported experiments to generate single photons (see, e.g., review articles \cite{scheel2009,eisaman2011,takeuchi2014,senellart2017,sinha2019} and references therein).

In this work, we ask \emph{how close} to a single photon we \emph{in principle} can come. The key requirement for states realizable on demand is causality: they cannot be measured before they are produced, meaning that the state must be indistinguishable from vacuum before the information from the source has had time to propagate to the observation point. We will also comment on situations where the state is generated with postselection instead, where the requirements are different. 

The maximum fidelity between a state realizable on demand and a single-photon state, will turn out to be dependent on the desired pulse form of the photon: Ultrashort pulses, with a pulse length of the order of a single cycle, or pulses with rapidly varying envelopes give a low fidelity. Similarly, the fidelity is low if the pulse is close to the onset and must be truncated. On the other hand, for pulses with slowly varying envelopes compared to the cycles, e.g., quasi-monochromatic Gaussian pulses, the realizable fidelity can be extremely close to unity. 

Bounds for single-photon approximations are clearly interesting from a theoretical point of view. First, they characterize the set of realizable states that approximate single photons. In addition, the analysis leads to a family of strictly localized states that are close to single photons. As the envelope width of the pulse increases, so does the fidelity. 

The bounds also have important implications for experimental quantum optics. Ultrashort pulses at the quantum level are now being generated in a number of schemes. As technology improves, it appears realistic to approach even single-cycle single-photon pulses \cite{su2016}. For ultrashort or rapidly varying pulses, the bounds are useful for estimating the best possible realizable single-photon state. 

In the Weisskopf-Wigner theory, an initially excited atom decays to its ground state under the production of a single photon \cite{weisskopf1930,scully1997}, in apparent contradiction to our results. However, the Weisskopf-Wigner theory is based on the rotating wave approximation, where antiresonant terms in the Hamiltonian are neglected. Without this approximation, the unitary evolution operator given by the Dyson series also produces small probability amplitudes for more than one photon. 

The rotating wave approximation also destroys causality, which is usually restored by a second approximation where the spectral integral is extended to negative frequencies. Similar approximations were done in the so-called Fermi problem \cite{fermi1932,hegerfeldt1994,buchholz1994}, where an initially excited atom couples via the electromagnetic field to a distant atom initially in the ground state. These problems are conveniently treated in the Heisenberg picture. Causality is then seen directly from the electric field operator, which becomes a superposition of the source-free operator and a sourced retarded-field operator \cite{milonni1995}, in complete analogy with classical electrodynamics.

\begin{figure}[tb]
    \includegraphics{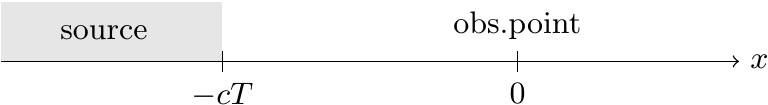}
    \caption{\label{fig:setup}A source is located in the region $x < -cT$, and we consider field observables at the observation point $x = 0$. The source is turned on at $t = -T$, which means that the observables must be unchanged for $t < 0$.}
\end{figure}

\emph{Setup and assumptions.} The setup in our analysis (see \fref{fig:setup}) consists of a plane-wave source, located at $x < -cT$, where $c$ is the vacuum light speed and $T > 0$ is a constant. For $t < -T$, the electromagnetic field is assumed to be in the vacuum state $\ket{0}$. The source is turned on at $t = -T$, off at $t = -T/2$, and we consider field observables at the observation point $x = 0$ after the source is done. 

We let $E(t)$ always denote the \emph{free} time evolution of the $E$-field at the observation point. In the interaction picture, the $E$-field retains the time dependence of the free theory, $E(t)$, and the states evolve according to the interaction-picture time-evolution operator $U$. This operator acts on the total Hilbert space of the electromagnetic field and the source, leading in general to a mixed optical state. 

Alternatively, we can use the Heisenberg picture, in which the full time evolution is included in the $E$-field operator, while the states remain unchanged. As shown in \cite{milonni1995}, the $E$-field then takes the form of the free-field part and a retarded-field part. By causality, for $t < 0$ the retarded field cannot yet have reached the observation point at $x = 0$, and thus the $E$-field there has the regular, free-field time dependence $E(t)$. This means that $U$ must leave $E(t)$ unchanged for $t < 0$:
\begin{equation}\label{eq:u_op_fund}
    U^\dagger E(t) U = E(t); \tbuff t < 0.
\end{equation}

Returning to the interaction picture, the optical state produced by the source is therefore given by applying a unitary operator $U$ satisfying \eqref{eq:u_op_fund} to the initial state $\ket{0} \otimes \ket{s}$ and then tracing out the source. Here, $\ket{s}$ is some arbitrary initial state of the source. 

In particular, the resulting optical state must therefore for $t < 0$ give expectation values of local field observables equal to that for vacuum. For example, $\exv{E^2(t)}$ and $\exv{\norder{E^2(t)}}$, where $\norder{}$ denotes normal order, will equal the corresponding expectation values for vacuum for $t < 0$. By definition \cite{knight1961,licht1963}, this means that the state is strictly localized to the region $t \geq 0$. In the \supplementary{}, we give a characterization of which field observables are local, and we show an alternative argument for requiring \eqref{eq:u_op_fund} based on physically measurable expectation values.

We consider plane-wave modes in the $+x$-direction with a single polarization \cite{blow1990}, expressing all quantities as a function of frequency (see, e.g., \cite{loudon2000}, Ch. 6). The operator $E(t)$ can then be decomposed into its positive- and negative-frequency parts \cite{loudon2000,cohen-tannoudji1997}:
\begin{equation}\label{eq:E_field}
    E(t) = E^+(t) + E^-(t),
\end{equation}
where
\begin{equation}\label{eq:E_pos}
    E^+(t) = \int_0^\infty \dd \omega \mathcal{E}(\omega) a(\omega) e^{-i \omega t}
\end{equation}
and $E^-(t) = \lp[E^+(t) \rp]^\dagger$. For later convenience, we write $\mathcal{E}(\omega) = K \sqrt{-i\omega}$, where $K > 0$ is a constant, absorbing the additional phase factor into $a(\omega)$. Moreover, $a(\omega)$ is the usual annihilation operator satisfying
\begin{equation}\label{eq:a_w_comm}
    \comm{a(\omega)}{a^\dagger(\omega')} = \delta(\omega - \omega').
\end{equation}

\emph{Pulse-mode formalism.} Similar to the treatment in \cite{tatarskii1990}, we introduce a complete, orthonormal set of pulse modes $\lp\{\xi_n(\omega) \rp\}_n$ spanning the function space $L^2(0, \infty)$:
\begin{subequations}
\begin{align}
    \int_0^\infty \dd \omega \xi_n^*(\omega) \xi_m(\omega) &= \delta_{nm}, \label{eq:xi_orthog} \\
    \sum_n \xi_n^*(\omega) \xi_n(\omega') &= \delta(\omega - \omega').
\end{align}
\end{subequations}
We can then define pulse-mode ladder operators by
\begin{equation}\label{eq:a_n_def}
    a_n^\dagger = \int_0^\infty \dd \omega \xi_n(\omega) a^\dagger(\omega).
\end{equation}
From \eqref{eq:a_w_comm} and \eqref{eq:xi_orthog} they satisfy
\begin{equation}\label{eq:a_n_comm}
    \comm{a_n}{a_m^\dagger} = \delta_{nm}.
\end{equation}
Using the complete set $\lp\{\xi_n(\omega) \rp\}_n$, we can now write
\begin{equation}\label{eq:E_pos_n}
    E^+(t) = \sum_n E_n(t) a_n,
\end{equation}
where
\begin{equation}\label{eq:E_n_def}
    E_n(t) = \int_0^\infty \dd \omega \mathcal{E}(\omega) \xi_n(\omega) e^{-i\omega t}.
\end{equation}

As an example, we can calculate the time dependence of the energy density of a single photon. The energy density operator minus the vacuum contribution is proportional to
\begin{equation}\label{eq:norder_E2}
    \norder{E^2(t)} = {E^+(t)}^2 + {E^-(t)}^2 + 2 E^-(t) E^+(t).
\end{equation}
The expectation value of \eqref{eq:norder_E2} for a single-photon state $\ket{1_1} = a_1^\dagger \ket{0}$ associated with some pulse mode $\xi_1(\omega)$ is easily obtained using \eqref{eq:a_n_comm}:
\begin{equation}\label{eq:exp_norder_E2}
    \bra{1_1} \norder{E^2(t)} \ket{1_1} = 2 \abs{E_1(t)}^2.
\end{equation}
According to \eqref{eq:E_n_def}, $E_1(t)$ contains only positive frequencies and is according to the Paley-Wiener criterion \cite{paley1934,papoulis1962} therefore nonzero (almost) everywhere. Since the expectation value of \eqref{eq:norder_E2} for vacuum is 0, and since \eqref{eq:exp_norder_E2} is nonzero for $t < 0$, single photons cannot be strictly localized \cite{knight1961,bialynicki-birula1998}.

On the other hand, coherent states $\ket{\alpha_{1}} = D_1(\alpha) \ket{0}$, $\alpha \in \complexs$ may be strictly localized \cite{bialynicki-birula1998}, since the displacement operator $D_1(\alpha) = \exp \lp( \alpha a_1^\dagger - \alpha^* a_1 \rp)$ is unitary and \footnote{This transformation can be found using the formula $e^{\gamma A}B e^{-\gamma A} = B + \gamma \comm{A}{B} + \frac{\gamma^2}{2!} \comm{A}{\comm{A}{B}} + \dotsb$}:
\begin{equation}\label{eq:coh_E_transf}
\begin{aligned}
    D_1^\dagger(\alpha) E(t) D_1(\alpha) &= E(t) + \alpha E_1(t) + \alpha^* E_1^*(t) \\
    &= E(t) + \int_{-\infty}^\infty \dd \omega F(\omega) e^{- i \omega t},
\end{aligned}
\end{equation}
where
\begin{equation}\label{eq:coh_F_def}
    F(\omega) =
    \begin{cases}
        \alpha \mathcal{E}(\omega) \xi_1(\omega); & \omega > 0, \\
        \alpha^* \mathcal{E}^*(-\omega) \xi_1^*(-\omega); & \omega < 0.
    \end{cases}
\end{equation}
The expected energy density is
\begin{equation}\label{eq:coh_E2_expv}
     \qexv{\alpha_1}{\norder{E^2(t)}} = \lp( \int_{-\infty}^\infty \dd \omega F(\omega) e^{- i \omega t} \rp)^2.
\end{equation}
Since the Fourier integral of $F(\omega)$ contains both positive and negative frequencies, the transformed $E(t)$ in \eqref{eq:coh_E_transf} can be made to satisfy \eqref{eq:u_op_fund} by a suitable spectrum $\xi_1(\omega)$, exactly as for classical pulses. In particular, there is a family of spectra that satisfy $F(\omega) = F^*(-\omega)$ and simultaneously have vanishing inverse Fourier transforms for $t < 0$.

\emph{Single-photon approximation.} The coherent state is not close to a single-photon state for any value of the parameter $\alpha$. Other localized states have also been suggested in the context of quantum field theory \cite{knight1961,licht1963}; however, it is unclear if these can be made close to single-particle states. 

\begin{figure}[tb]
    \includegraphics{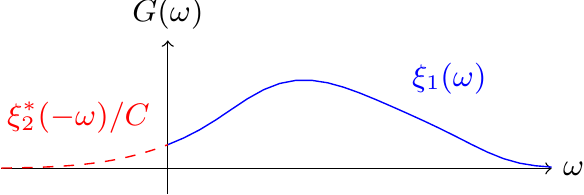}
    \caption{\label{fig:G}The absolute value of a typical $G(\omega)$.}
\end{figure}

We will now consider the possibility of having states that are strictly localized to $t\geq 0$, and yet close to a single photon. The intuition behind avoiding the infinite tails of one pulse mode is to compensate the negative-time components with small terms of another mode. A condition that will turn out to be sufficient for achieving this is to find two pulse modes $\xi_1(\omega)$ and $\xi_2(\omega)$, such that
\begin{equation}\label{eq:cond_E2_E1}
    E_2(t) = - C E_1^*(t); \tbuff t < 0, 
\end{equation}
where $C$ is a real constant. Note that the condition is only supposed to be valid for $t < 0$. To see how \eqref{eq:cond_E2_E1} can be achieved, define
\begin{equation}\label{eq:f_def}
    f(t) = E_1(t) + E_2^*(t) / C.
\end{equation}
From \eqref{eq:E_n_def} we find that
\begin{equation}\label{eq:f_G_Fourier}
    f(t) = \int_{-\infty}^\infty \dd \omega \mathcal{E}(\omega) G(\omega) e^{-i\omega t},
\end{equation}
where
\begin{equation}\label{eq:G_def}
    G(\omega) = 
    \begin{cases}
        \xi_1(\omega); & \omega > 0, \\
        \xi_2^*(-\omega) / C; & \omega < 0.
    \end{cases}
\end{equation}
To satisfy \eqref{eq:cond_E2_E1}, we must choose $G(\omega)$ such that the inverse transform $f(t)$ vanishes for $t < 0$ \footnote{The function $f(t)$ is not the inverse Fourier transform of $G(\omega)$, but the inverse Fourier transform of $\mathcal{E}(\omega) G(\omega)$. However, $\mathcal{E}(\omega) = K \sqrt{-i\omega}$ is an analytic function in the upper half-plane of complex frequency $\omega$, since the branch cut can be taken elsewhere. Therefore, if the inverse Fourier transform of $G(\omega)$ vanishes for $t < 0$, so does $f(t)$. Here it is assumed that we have picked a $G(\omega)$ that tends sufficiently fast to zero as $\omega \to \infty$ in the upper half-plane. If the goal instead is localization on both sides, i.e., $f(t) = 0$ outside an interval $0 \leq t \leq \mathcal T$, one must start with a suitable $f(t)$ and calculate the required $G(\omega)$. It can be shown that the procedure for two-sided localization is also consistent with the modification of $G(\omega)$ in the \supplementary{}.}. This function $G(\omega)$ must necessarily be nonzero for (almost) all $\omega$. Typically, we pick a function with its main weight for positive frequencies (see \fref{fig:G}). We then multiply $G(\omega)$ by a real constant such that its restriction to positive frequencies, $\xi_1(\omega)$, is normalized. Finally, $\xi_2(\omega)$ is found from $G(\omega)$ after determining $C$ such that $\xi_2(\omega)$ also gets normalized. There is a small complication: We must ensure that the resulting pulse modes $\xi_1(\omega)$ and $\xi_2(\omega)$ are orthogonal. In the \supplementary{}, it is shown that this can always be done easily by a very small modification of $G(\omega)$, which can be neglected in the following analysis.

Define $\eta$ as the fraction of the squared norm of $G(\omega)$ located for negative frequencies. It follows that 
\begin{equation}\label{eq:C_r}
    C^2 = (1 - \eta) / \eta.
\end{equation}

We will now look for a unitary operator $U$ acting on the source and pulse modes 1 and 2, to create a state that is localized to $t \geq 0$. With condition \eqref{eq:cond_E2_E1}, the electric field operator for $t < 0$ can be written
\begin{align}\label{eq:e_field_C_cond}
    E(t) &= E_1(t) \lp(a_1 - Ca_2^\dagger \rp) + \sum_{n \geq 3} E_n(t) a_n + \hc
\end{align}
To achieve \eqref{eq:u_op_fund}, it is therefore sufficient that
\begin{equation}\label{eq:comm_U_a1_a2}
    \comm{U}{a_1 - C a_2^\dagger} = 0.
\end{equation}
Consider a source Hilbert space with two basis states $\ket{\text{g}}, \ket{\text{e}}$ and source excitation / de-excitation operators $\sigma_+ = \ket{\text{e}}\bra{\text{g}}$, $\sigma_- = \ket{\text{g}}\bra{\text{e}}$. Define the unitary operator
\begin{equation}\label{eq:U_S_def}
    U_\text{S} = A_1^\dagger \sigma_- + A_1 \sigma_+ + \ket{0_1}\bra{0_1} \otimes \ket{\text{g}}\bra{\text{g}},
\end{equation}
with
\begin{equation}\label{eq:A_1_def}
    A_1^\dagger = a_1^\dagger \frac{1}{\sqrt{a_1 a_1^\dagger}} = \sum_n \ket{{n + 1}_1}\bra{n_1},
\end{equation}
where $\ket{n_k} = {a_k^\dagger}^n \ket{0_k} / \sqrt{n!}$. Further, for $\tanh\gamma = 1 / C$, define
\begin{equation}\label{eq:S_def}
    S = e^{\gamma a_1 a_2 - \gamma a_1^\dagger a_2^\dagger},
\end{equation}
which is also unitary and describes the transformation \cite{Note1,leonhardt2003}:
\begin{equation}\label{eq:S_transf}
    S \lp(a_1 - C a_2^\dagger \rp) S^\dagger = - a_2^\dagger \sqrt{C^2 - 1}.
\end{equation}
Since $U_{\text{S}}$ commutes with \eqref{eq:S_transf}, it follows that the unitary operator $U \equiv S^\dagger U_{\text{S}} S$ satisfies \eqref{eq:comm_U_a1_a2}. Tracing out the source space from $U \ket{0} \otimes \ket{\text{e}}$ means that
\begin{equation}\label{eq:r_state_def}
    \ket{\eta_{1,2}} \equiv S^\dagger A_1^\dagger S \ket{0}
\end{equation}
is a strictly localized state, with the form
\begin{equation}\label{eq:r_state_expansion}
    \ket{\eta_{1,2}} = c_1 \ket{1_1 \kbuff 0_2} + c_2 \ket{2_1 \kbuff 1_2} + c_3 \ket{3_1 \kbuff 2_2} + \dotsb,
\end{equation}
for coefficients $c_1, c_2, \dotsc$ In practice we will often have $\eta \ll 1$, and we can then find the fidelity between $\ket{\eta_{1,2}}$ and the single-photon state $\ket{1_1 \kbuff 0_2}$ by \eqref{eq:C_r} and expanding \eqref{eq:r_state_def} for small $\eta$:
\begin{equation}\label{eq:r_state_fidelity}
    F \equiv \abs{\braket{1_1 \kbuff 0_2}{\eta_{1,2}}} = 1 - \frac{3 - 2 \sqrt{2}}{2} \eta + \order{\eta^2} \approx 1 - 0.09 \eta.
\end{equation}

We have arrived at the first of our two main results. Equation \eqref{eq:r_state_def} expresses a strictly localized state that is close to a single photon as measured by the fidelity \eqref{eq:r_state_fidelity}. For this state, the expectation value of any local observable is for $t < 0$ the same as that for vacuum. 

Additionally, instead of restricting the pulse to $t \geq 0$, we can just as easily choose $0 \leq t \leq \mathcal{T}$, for some constant $\mathcal{T} > 0$. Using the same procedure, we can then find a state that is strictly localized to a bounded interval \cite{Note2} while being close to a single photon. In particular, its energy density will be equal to that for vacuum everywhere outside this interval.

\emph{Bounds.} Finally, we aim to find inequalities that constrain how close to a desired optical state we can come with a strictly localized state. One possible choice for the desired state is simply a single photon with spectrum $\xi(\omega)$. Such a state will however necessarily give a tail for negative $t$, and is therefore not always the most suitable representation of the desired photon pulse form. We therefore consider the desired state to instead be a single photon in some specified, positive-time pulse $g(t)$:
\begin{equation}\label{eq:1g_state_def}
    \ket{1_g} = \int_{0}^\infty \dd t g(t) a^\dagger(t) \ket{0}; \tbuff \int_{0}^\infty \dd t \abs{g(t)}^2 = 1,
\end{equation}
where $g(t) = 0$ for $t < 0$. Here, $a^\dagger(t)$ is a time-domain creation operator satisfying $\comm{a(t)}{a^\dagger(t')} = \delta(t - t')$ \cite{loudon2000}. At the same time, $a^\dagger(t)$ is the Fourier transform of $a^\dagger(\omega)$, and the required negative-frequency modes must be seen as an artificial construction to be able to express the desired state. Provided we extend $E(t)$ with these negative frequencies as well, $\ket{1_g}$ is an artificial, single-photon state that is localized to $t \geq 0$. 

Any physical state $\ket{\psi}$ contains a superposition of (products of) ladder operators $a^\dagger(\omega)$ only for $\omega > 0$, acting on the vacuum state. Therefore, the maximum fidelity satisfies
\begin{equation}\label{eq:F_max_def}
    F_{\text{max}} \equiv \max_{\ket{\psi}} \abs{\braket{\psi}{1_g}} = \max_{\ket{\psi}} \abs{\bra{\psi} \int_0^\infty \dd \omega G(\omega) a^\dagger(\omega) \ket{0}},
\end{equation}
where $G(\omega)$ is the Fourier transform of $g(t)$. Using the Cauchy-Schwarz inequality, we obtain an upper bound
\begin{equation}\label{eq:F_max_upper}
    F_{\text{max}}^2 \leq \int_{0}^\infty \dd \omega \abs{G(\omega)}^2 = 1 - \eta.
\end{equation}  

A lower bound for $F_{\text{max}}$ can now be found by construction, using the results above. Given a desired state \eqref{eq:1g_state_def}, we first define a corresponding single-photon state with only positive frequencies:
\begin{equation}\label{eq:1g_pos_def}
    \ket{1_g^+} = \frac{1}{\sqrt{1 - \eta}} \int_{0}^{\infty} \dd \omega G(\omega) a^\dagger(\omega) \ket{0}.
\end{equation}
By the identification \eqref{eq:G_def}, the state $\ket{1_g^+}$ plays the role of $\ket{1_1 \kbuff 0_2}$ in \eqref{eq:r_state_fidelity}. Thus we can approximate $\ket{1_g^+}$ by a localized state $\ket{\eta_{1,2}}$, with fidelity $F = \abs{\braket{1_g^+}{\eta_{1,2}}} \approx 1 - 0.09\eta$. This gives
\begin{equation}\label{eq:F_max_lower}
    F_{\text{max}} \geq \abs{\braket{1_g}{\eta_{1,2}}} = \abs{\braket{1_g^{\vphantom{+}}}{1_g^+} \braket{1_g^+}{\eta_{1,2}}} = \sqrt{1 - \eta} F.
\end{equation}
We conclude that
\begin{equation}\label{eq:F_max_bounds}
    1 - 0.59\eta \approx F \sqrt{1 - \eta} \leq F_{\text{max}} \leq \sqrt{1 - \eta} \approx 1 - 0.5\eta,
\end{equation}
where the approximations are valid for $\eta \ll 1$. The bounds \eqref{eq:F_max_bounds} represent our second main result, constraining the fidelity between states generated locally on demand, and a desired single photon in a given pulse $g(t)$. 

It is clear that the bounds also apply to mixed states $\rho = \sum_i p_i \ket{\psi_i} \bra{\psi_i}$, where $p_i$ are probabilities and $\ket{\psi_i}$ are corresponding states. The fidelity is then given by $F^2_{\text{max}} = \max_\rho \bra{1_g} \rho \ket{1_g} = \max_{\ket{\psi}} \abs{\braket{\psi}{1_g}}^2$.

Similarly, if the source is operated with postselection, the upper bound for $F_{\text{max}}$ is still valid because of the artificial negative frequencies of the desired state \eqref{eq:1g_state_def}. The lower bound is on the other hand not valid, since we in light of the Reeh-Schlieder theorem \cite{reeh1961,licht1963,haag1996} can approximate a single-photon state \eqref{eq:1g_pos_def} arbitrarily well by a selective local operation. The lower bound is therefore simply equal to the upper bound, however with the caveat that any given trial has a limited probability of being postselected. 

\begin{figure}[tb]
    \includegraphics{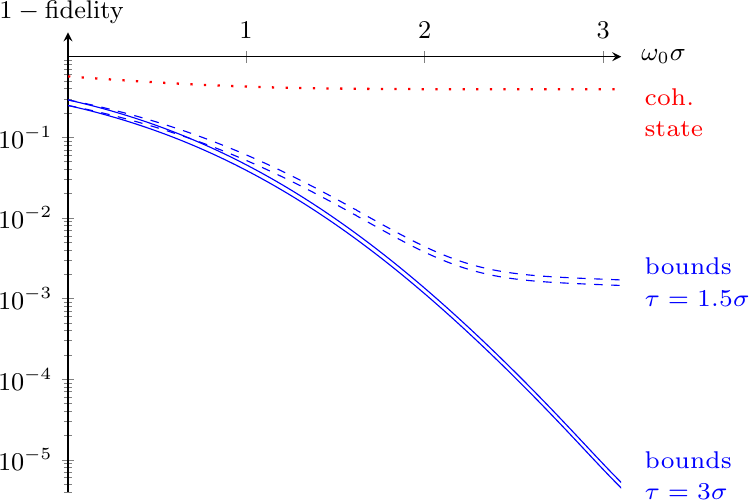}
    \caption{\label{fig:fidelity} $1 - F_{\text{max}}$, where $F_{\text{max}}$ is the maximum fidelity when the desired pulse envelope is a truncated Gaussian with carrier $\omega_0$, duration $\sigma$, and delay $\tau$: $g(t) \propto u(t) e^{-(t-\tau)^2/2\sigma^2} e^{-i \omega_0 t}$. The two solid lines ($\tau=3\sigma$) show the upper and lower bounds as given by \eqref{eq:F_max_bounds} to first order in $\eta$. Similarly, the dashed lines show the bounds when $\tau=1.5\sigma$. The differences between the next orders in $\eta$ suggest that the approximation is good for $\omega_0 \sigma \gtrsim 0.3$. The dotted line shows $1 - \text{fidelity}$ of the truncated Gaussian with a coherent state with $\alpha = 1$, which approaches $1/\sqrt{e}$ as $\omega_0 \sigma \to \infty$ (the fidelity is almost identical for $\tau=1.5\sigma$ and $\tau=3\sigma$).}
\end{figure}

As an example, consider a target pulse that is a truncated Gaussian with carrier frequency $\omega_0$, duration $\sigma$, and delay $\tau$:
\begin{equation}
    g(t) \propto u(t) e^{-(t-\tau)^2/2\sigma^2} e^{-i\omega_0 t}, 
\end{equation}
where $u(t)$ is the Heaviside function. Truncation of the leading tail is required \footnote{Alternatively, we can make the pulse continuous by a short interpolation region after $t=0$, moving the discontinuity to the derivative instead. In any case, there must be some non-analytic point representing the information from the person controlling the source, which must be present for any on-demand source.} because causality necessitates $g(t) = 0$ for $t < 0$ from \eqref{eq:1g_state_def}. By calculating the spectrum, one finds $\eta$ and the values for the bounds \eqref{eq:F_max_bounds} (see \fref{fig:fidelity}). Clearly the fidelity is much higher than that with a coherent state. For single-cycle pulses, $\omega_0 \sigma \lesssim 1$, the fidelity is low since the pulse contains a significant amount of negative frequencies, giving a large $\eta$. As $\omega_0 \sigma$ increases, the fidelity increases rapidly up to a point where the truncation dominates the negative-frequency content and the fidelity remains roughly constant. By increasing the pulse delay $\tau$, there is less truncation and the fidelity can be made arbitrarily close to 1 by a sufficiently large $\sigma$. 

We can also consider the limiting case of a Gaussian pulse without truncation by taking $\omega_0 \tau \rightarrow \infty$, corresponding to delaying the pulse infinitely long. This case can be treated analytically, giving
\begin{equation}\label{eq:r_limit_inf_delay}
	\eta = \frac{1}{2}\lp[ 1 - \erf(\omega_0\sigma) \rp],
\end{equation}
where $\erf(\cdot)$ is the error function. The result is practically indistinguishable from the case $\tau=3\sigma$ (corresponding to a large, finite delay) in \fref{fig:fidelity}. This regime is a good approximation for most on-demand single-photon sources today, where we typically have an externally pumped system that emits a photon by relaxation \cite{flagg2012,khramtsov2017}. The delay $\omega_0 \tau$ is usually large, either from a long spontaneous emission time or from the time taken for triggering a controlled emission. However, if such devices can be operated with very short pulse times $\omega_0 \sigma$, eq. \eqref{eq:r_limit_inf_delay} shows that the fidelity is still limited even for arbitrarily long relaxation times. We emphasize that the bounds for the fidelity are a fundamental consequence of the field itself, and are therefore valid for all possible sources. 

In conclusion, we have constrained the optimal fidelity of locally generated optical states that approximate single-photon pulses. Perfect single photons are in principle unrealizable on demand, but the optimal fidelity increases rapidly as the pulse envelope becomes more slowly varying. In addition to implications for ultrafast optics, our results demonstrate optical states that are strictly localized yet close to single photons.

\begin{acknowledgments}
\emph{Acknowledgments.} We thank Jon Magne Leinaas and Joakim Bergli for helpful suggestions and comments. 
\end{acknowledgments}


%

\appendix
\section*{\supplementary{}}
\setcounter{equation}{0}
\renewcommand*{\theequation}{S\arabic{equation}}

\subsection{Local field observables and localized states}
In this section, we provide additional justification for the on-demand realizability condition (1) in the main document and discuss which field observables are considered local.

Let $\hilbertspace_{\text{em}}$ be the optical state space and $\hilbertspace_{\text{src}}$ the state space of the source. From the main document, we recall that a state $\rho$ produced by a localized, on-demand source is given by
\begin{equation}\label{eq:realizable_state}
    \rho = \tr_{\text{src}} \lp(U \ket{0 \kbuff s} \bra{0 \kbuff s} U^\dagger \rp),
\end{equation}
where the unitary operator $U$ leaves $E(t)$ unchanged for time $t < 0$:
\begin{equation}\label{eq:u_op_fund_app}
    U^\dagger E(t) U = E(t); \tbuff t < 0.
\end{equation}
Here, $\ket{0 \kbuff s} = \ket{0} \otimes \ket{s}$ is the product state of electromagnetic vacuum in all modes and an arbitrary initial source state $\ket{s} \in \hilbertspace_{\text{src}}$, $E(t)$ is the free electric field operator and the trace is over the source degrees of freedom. This condition was argued based on the causality analysis in \cite{milonni1995}, as the retarded-field contribution from the source has not yet reached the observation point for $t < 0$. 

However, \eqref{eq:u_op_fund_app} is an operator equation, and causality-based arguments are more naturally formulated in terms of physically measurable quantities. In the following, we give an alternative condition based on this line of reasoning and argue that it is equivalent to \eqref{eq:u_op_fund_app}. Another discussion of \eqref{eq:u_op_fund_app} and its relation to strict localization can be found in \cite{licht1963}.

Let $\ket{\Psi} \in \hilbertspace_{\textnormal{em}} \otimes \hilbertspace_{\textnormal{src}}$ be an initial state of the total system, electromagnetic and source. Relativistic causality tells us that the effect of the source, $U$, should not be visible outside the light cone, meaning that any physically measurable expectation value, such as $\exv{E(t)}$, remains unchanged for $t < 0$:
\begin{equation}\label{eq:equal_expval}
    \bra{\Psi} U^\dagger E(t) U \ket{\Psi} = \bra{\Psi} E(t) \ket{\Psi}; \tbuff t < 0.
\end{equation}
This must hold true for all initial states $\ket{\Psi}$. To see the equivalence with condition \eqref{eq:u_op_fund_app}, we first note that \eqref{eq:u_op_fund_app} trivially implies \eqref{eq:equal_expval}. Conversely, \eqref{eq:equal_expval} forces the spectrum of the operator $A = U^\dagger E(t) U - E(t)$ for $t < 0$ to contain only the single point 0, which gives $A = 0$ by the spectral theorem.

Of course, we should further require \eqref{eq:equal_expval} to hold not only for $E(t)$, but for \emph{all} local observables, since all possible measurements must remain unchanged outside the source's light cone. However, since \eqref{eq:equal_expval} forces the operator $E(t)$ itself to be unchanged for $t < 0$, we can show that this automatically ensures that all local operators remain unchanged here as well. 

To see this, we consider the set of local operators \cite{knight1961}, which are superpositions and products of the field evaluated in the measurement region $t < 0$. Let the state be given by \eqref{eq:realizable_state} and \eqref{eq:u_op_fund_app}. Starting with the expectation value of the field, we obtain for $t < 0$:
\begin{equation}\label{eq:exv_E}
\begin{aligned}
    \exv{E(t)} &= \qexv{0 \kbuff s}{U^\dagger E(t) U} = \qexv{0 \kbuff s}{E(t)} \\
    &= \qexv{0}{E(t)},
\end{aligned}
\end{equation}
so $\exv{E(t)}$ is equal to the corresponding vacuum value. Consider next the expectation value of a product:
\begin{equation}\label{eq:exv_E_prod}
\begin{aligned}
    \exv{E(t_1) \dotsm E(t_n)} &= \qexv{0 \kbuff s}{U^\dagger E(t_1) \dotsm E(t_n) U} \\
    &= \qexv{0}{E(t_1) \dotsm E(t_n)},
\end{aligned}
\end{equation}
where $t_1, \dotsc, t_n < 0$. Superpositions of field expressions follow trivially. 

Other operators of interest are normal-ordered products of fields. It is these we in practice typically measure, since non-normal-ordered products may yield infinite expectation values. Note however that normal ordering is not automatically guaranteed to give a local operator. In the case of \emph{free} fields it is valid, since normal-ordered products are then always built up of other local operators and constants, e.g., 
\begin{equation}\label{eq:exv_E_norder_2}
\begin{aligned}
    \exv{\norder{E(t_1)E(t_2)}} &= \exv{E(t_1)E(t_2)} - \comm{E^+(t_1)}{E^-(t_2)} \\ 
    &= \exv{E(t_1)E(t_2)} - \qexv{0}{E(t_1)E(t_2)} \\
    &= 0; \tbuff t_1, t_2 < 0,
\end{aligned}
\end{equation}
where $E^\pm(t)$ are the positive- and negative-frequency components of $E(t)$ and $\norder{}$ denotes normal order. In the last equality, we have used \eqref{eq:exv_E_prod}. As a special case, in the limit $t_2 \to t_1$, we find that the expected energy density $\exv{\norder{E(t_1)^2}}$ vanishes for $t_1 < 0$. 

For normal-ordered $n$ factors of fields we get that 
\begin{equation}\label{eq:exv_E_norder_n}
    \exv{\norder{E(t_1) \dotsm E(t_n)}} = 0; \tbuff t_1, \dotsc, t_n < 0.
\end{equation}
To prove \eqref{eq:exv_E_norder_n}, we use a version of Wick's theorem without time ordering, which is obtained along the lines of the conventional proof (see, e.g., \cite{peskin1995}, Ch. 4). For example for $n = 4$, we have
\begin{align}\label{eq:wick_4}
    & \exv{E(t_1)E(t_2)E(t_3)E(t_4)} = \exv{\norder{E(t_1)E(t_2)E(t_3)E(t_4)}} \nonumber \\
    &+ \contraction{}{E(t_1)}{}{E(t_2)} E(t_1)E(t_2) \exv{\norder{E(t_3)E(t_4)}}
    + \contraction{}{E(t_1)}{}{E(t_3)} E(t_1)E(t_3) \exv{\norder{E(t_2)E(t_4)}} \nonumber \\
    &+ \contraction{}{E(t_1)}{}{E(t_4)} E(t_1)E(t_4) \exv{\norder{E(t_2)E(t_3)}}
    + \contraction{}{E(t_2)}{}{E(t_3)} E(t_2)E(t_3) \exv{\norder{E(t_1)E(t_4)}} \nonumber \\
    &+ \contraction{}{E(t_2)}{}{E(t_4)} E(t_2)E(t_4) \exv{\norder{E(t_1)E(t_3)}}
    + \contraction{}{E(t_3)}{}{E(t_4)} E(t_3)E(t_4) \exv{\norder{E(t_1)E(t_2)}} \nonumber \\
    &+ \contraction{}{E(t_1)}{}{E(t_2)}\contraction{E(t_1)E(t_2)}{E(t_3)}{}{E(t_4)} E(t_1)E(t_2)E(t_3)E(t_4)
    + \contraction{}{E(t_1)}{E(t_2)}{E(t_3)}\contraction[2ex]{E(t_1)}{E(t_2)}{E(t_3)}{E(t_4)} E(t_1)E(t_2)E(t_3)E(t_4) \nonumber\\
    &+ \contraction[2ex]{}{E(t_1)}{E(t_2)E(t_3)}{E(t_4)}\contraction{E(t_1)}{E(t_2)}{}{E(t_3)} E(t_1)E(t_2)E(t_3)E(t_4),
\end{align}
where the contractions are defined without time ordering:
\begin{equation}\label{eq:contraction}
    \contraction{}{E(t_1)}{}{E(t_2)} E(t_1)E(t_2) = \bra{0} E(t_1) E(t_2) \ket{0}.
\end{equation}
By using \eqref{eq:wick_4} twice, once for $\exv{E(t_1) \dotsm E(t_4)}$ and once for $\qexv{0}{E(t_1) \dotsm E(t_4)}$, we find:
\begin{equation}\label{eq:exv_E_norder_4}
\begin{aligned}
    0 &= \exv{E(t_1) \dotsm E(t_4)} - \qexv{0}{E(t_1) \dotsm E(t_4)} \\
    &= \exv{\norder{E(t_1)E(t_2)E(t_3)E(t_4)}},
\end{aligned}
\end{equation}
where we have used \eqref{eq:exv_E_prod} and \eqref{eq:exv_E_norder_2}.

Of course, all these results also apply to the magnetic field, field derivatives, and mixed products of fields and field derivatives. For example, from \eqref{eq:exv_E_norder_2} one obtains $\exv{\norder{E(t_1)E'(t_2)}} = 0$, as
\begin{equation}\label{eq:exv_E_Eder}
    E(t_1) E'(t_2) = \lim_{\Delta t_2 \to 0} \frac{E(t_1) E(t_2 + \Delta t_2) - E(t_1) E(t_2)}{\Delta t_2}.
\end{equation}

On the other hand, nonlocal field observables will not necessarily give the vacuum result for $t < 0$. One such observable is the conventional intensity operator $E^-(t) E^+(t)$ from Glauber's detection theory \cite{glauber1963,loudon2000}. Determining $E^\pm(t)$ from $E(t)$ involves the Hilbert transform, which is nonlocal in time; $E^\pm(t)$ for a fixed time $t$ is dependent on $E(t)$ for all $t$. Therefore, the expectation value of the intensity operator does not necessarily vanish completely for $t < 0$, even for states that are strictly localized to $t \geq 0$. This nonlocal property of the intensity operator is an artifact resulting from the use of the rotating wave approximation in its derivation \cite{bykov1989,milonni1995}, analogous to the quasi-monochromatic assumption in the derivation of the classical intensity $\abs{E}^2$.

Finally, note the distinction between local operators and local observables. As we have shown, \eqref{eq:u_op_fund_app} (or equivalently \eqref{eq:equal_expval}) ensures that all local operators are unchanged for $t < 0$. This encompasses all the local observables, which are the Hermitian local operators, but includes other operators as well. For example, the operator $E(t_1) E(t_2)$ is local and has an unchanged expectation value for $t_1, t_2 < 0$, but is not Hermitian and therefore not an observable. On the other hand, $E(t_1) E(t_2) + E(t_2) E(t_1)$, which also has an unchanged expectation value for negative times, is a local observable.

\subsection{Modification of \texorpdfstring{$\bm{G(\omega)}$}{G(w)}}
In the main document, we identify two pulse modes $\xi_1(\omega)$ and $\xi_2(\omega)$ from a function $G(\omega)$ using
\begin{equation}\label{eq:G_def_app}
    G(\omega) = 
    \begin{cases}
        \xi_1(\omega); & \omega > 0, \\
        \xi_2^*(-\omega) / C; & \omega < 0,
    \end{cases}
\end{equation}
where $C$ is some real constant. Since $\xi_1(\omega)$ and $\xi_2(\omega)$ are orthogonal, we must require
\begin{equation}\label{eq:G_orthog}
    \int_0^\infty \dd \omega G(\omega) G(-\omega) = 0.
\end{equation}
At the same time, $G(\omega)$ is a function whose inverse Fourier transform $g(t)$ vanishes for $t < 0$. For a given such function $G(\omega)$, the orthogonality relation \eqref{eq:G_orthog} is not necessarily satisfied. However, we can modify $G(\omega)$ such that the relation is satisfied. Define
\begin{subequations}\label{eq:G_prop_defs}
\begin{align}
    \norm{G}^2 &= \int_{-\infty}^\infty \dd \omega \abs{G(\omega)}^2, \label{eq:func_norm} \\
    \eta &= \frac{1}{\norm{G}^2} \int_{-\infty}^0 \dd \omega \abs{G(\omega)}^2, \label{eq:r_def} \\
    I &= \frac{1}{\norm{G}^2} \int_0^\infty \dd \omega G(\omega)G(-\omega). \label{eq_I_def}
\end{align}
\end{subequations}
Note that \eqref{eq:r_def} coincides with the definition of $\eta$ in the main document. Also, by redefining $G(\omega) \mapsto G^*(-\omega)$ if necessary, we can always assume that $0 \leq \eta < 1/2$. Here we have for simplicity excluded the limiting case $\eta = 1/2$. 

Let the modified function be
\begin{equation}\label{eq:G_tilde}
    \widetilde{G}(\omega) = G(\omega) - \beta G^*(-\omega),
\end{equation}
where $\beta$ is a complex constant. First, we note that the inverse Fourier transform of $\widetilde{G}(\omega)$ is $\widetilde{g}(t) = g(t) - \beta g^*(t)$, which clearly vanishes for $t < 0$. Hence, \eqref{eq:G_tilde} is a valid modification, and $\widetilde{G}(\omega)$ is non-vanishing since $\eta \neq 1/2$. Using $\widetilde{G}(\omega)$ rather than $G(\omega)$ in \eqref{eq:G_orthog}, we obtain
\begin{equation}\label{eq:G_tilde_orthog_cond}
    \beta^2 I^* - \beta + I = 0,
\end{equation}
with solution (choosing the minus sign):
\begin{equation}\label{eq:beta_sol}
    \beta = \frac{1}{2I^*} \lp(1 - J \rp),
\end{equation}
where
\begin{equation}\label{eq:J_def}
    J = \sqrt{1 - 4 \abs{I}^2}.
\end{equation}
The Cauchy-Schwarz inequality applied to \eqref{eq_I_def} yields
\begin{equation}\label{eq:I2_upper}
    \abs{I}^2 \leq (1 - \eta) \eta.
\end{equation}
Usually $\eta \ll 1$, which gives $\abs{I} \ll 1$ and $\beta \approx I$. 

Proceeding with the exact calculation, we define
\begin{equation}\label{eq:r_tilde_def}
    \widetilde{\eta} = \frac{1}{\bignorm{\widetilde{G}}^2} \int_{-\infty}^0 \dd \omega \bigabs{\widetilde{G}(\omega)}^2,
\end{equation}
and obtain after some lengthy, but trivial algebra that
\begin{equation}\label{eq:r_tilde_sol}
    \widetilde{\eta} - \eta = - \frac{1 - J}{2J} \lp(1 - 2\eta \rp).
\end{equation}
It follows that $\widetilde{\eta} \leq \eta$. In other words, the modification $G(\omega) \mapsto \widetilde{G}(\omega)$ actually improves $\eta$, in the sense that the tail for negative frequencies gets smaller. The improvement increases with the overlap between $G(\omega)$ and $G^*(-\omega)$. 

On the other hand, when the original $G(\omega)$ is a ``desired'' spectrum of some specified state
\begin{equation}\label{eq:1g_pos_def_app}
    \ket{1_g^+} = \int_0^\infty \dd \omega \frac{G(\omega)}{\norm{G}_+} a^\dagger(\omega) \ket{0},
\end{equation}
the modification leads of course to a worsening. Here, $\norm{ \cdot }_+$ is the function norm \eqref{eq:func_norm} restricted to positive frequencies and $a^\dagger(\omega)$ is the usual creation operator satisfying $\comm{a(\omega)}{a^\dagger(\omega')} = \delta(\omega - \omega')$. In the main document, we show how $\ket{1_g^+}$ can be approximated by a localized state $\ket{\eta_{1,2}}$, with fidelity
\begin{equation}\label{eq:r_state_fidelity_app}
    F = 1 - \frac{3 - 2\sqrt{2}}{2}\eta + \order{\eta^2}.
\end{equation}
This is however ignoring the modification $G(\omega) \mapsto \widetilde{G}(\omega)$, and so $F$ actually gives the fidelity of how close we can approximate the modified state
\begin{equation}\label{eq:1g_tilde_pos_def}
    \bigket{1_{\widetilde{g}}^+} = \int_0^\infty \dd \omega \frac{\widetilde{G}(\omega)}{\bignorm{\widetilde{G}}_+} a^\dagger(\omega) \ket{0}
\end{equation}
by a localized state, meaning that
\begin{equation}\label{eq:F_def_app}
    F = \bigabs{\bigbraket{1_{\widetilde{g}}^+}{\eta_{1,2}}}. 
\end{equation}
Here we are, as in the main document, neglecting $\eta \mapsto \widetilde{\eta}$, since it is an improvement (in order to make use of this result, we must make the additional, mild assumption that the desired spectrum $G(\omega)$ either satisfies $\eta < 1/2$ or that we are allowed to take $G(\omega) \mapsto G^*(-\omega)$ if it does not). We can find a lower bound for the fidelity $\abs{\braket{1_g^+}{\eta_{1,2}}}$ by converting to trace distance and using the triangle inequality:
\begin{equation}\label{eq:trace_dist_triangle}
\begin{aligned}
    &\sqrt{1 - \abs{\braket{1_g^+}{\eta_{1,2}}}^2} \\
    &\leq \sqrt{1 - \bigabs{\bigbraket{1_{\vphantom{\widetilde{g}}g}^+}{1_{\widetilde{g}}^+}}^2} +  \sqrt{1 - \bigabs{\bigbraket{1_{\widetilde{g}}^+ }{\eta_{1,2}}}^2}. 
\end{aligned}
\end{equation}
Using \eqref{eq:I2_upper}, we can find a lower bound for the inner product between $\ket{1_g^+}$ and $\bigket{1_{\widetilde{g}}^+}$:
\begin{equation}\label{eq:in_prod_G_G_tilde}
\begin{aligned}
    \bigabs{\bigbraket{1_{\vphantom{\widetilde{g}}g}^+}{1_{\widetilde{g}}^+}}^2 &= \frac{1}{\bignorm{G}^2_+ \bignorm{\widetilde{G}}^2_+} \abs{\int_{0}^\infty \dd \omega G(\omega) \widetilde{G}^*(\omega)}^2 \\
    &= \frac{\lp(1 + J - 2\eta \rp) \lp(1 + J \rp)}{4 J \lp(1 - \eta \rp)} \\
    &= 1 + \abs{I}^4 - \abs{I}^2 \lp(\eta + \eta^2 \rp) + \order{\eta^3} \\
    &\geq 1 - \eta^2 + \order{\eta^3}.
\end{aligned}
\end{equation}
By inserting \eqref{eq:r_state_fidelity_app} and \eqref{eq:in_prod_G_G_tilde} into \eqref{eq:trace_dist_triangle} and expanding for $\eta \ll 1$, we get that
\begin{equation}
    \abs{\braket{1_g^+}{\eta_{1,2}}} \geq 1 - \frac{3 - 2\sqrt{2}}{2} \eta + \order{\eta^{3/2}}.
\end{equation}
Hence, $F$ is also a lower bound for $\abs{\braket{1_g^+}{\eta_{1,2}}}$ to first order in $\eta$. 

In both cases, we can neglect the modification $G(\omega) \mapsto \widetilde{G}(\omega)$, as it leads either to an improvement or a small correction that can be ignored for sufficiently small $\eta$.

\onecolumngrid
\end{document}